%% file: 00-main.tex
\documentclass[acmsmall,screen]{acmart}
\usepackage[utf8]{inputenc}
\usepackage{multirow}
\usepackage{tabularx}
\usepackage{booktabs} 
\usepackage{hyperref}
\usepackage{amsmath}
\usepackage{verbatim}
\usepackage{siunitx}
\usepackage[shortlabels]{enumitem}
\usepackage{fancybox}
\usepackage{url}
\usepackage{graphicx}
\usepackage{setspace}
\usepackage{soul}
\usepackage[roman]{parnotes}
\usepackage{color}
\usepackage{colortbl}
\usepackage{listings}
\usepackage{subcaption}
\usepackage{algorithm, algpseudocode}
\algnewcommand\algorithmicto{\textbf{to}}
\algnewcommand\RETURN{\State \textbf{return} }
\usepackage{textcomp}
\usepackage{xcolor}
\usepackage[justification=centering]{caption}
\usepackage{pifont}

\begin{document}

\title{Beyond Self-learned Attention: Mitigating Attention Bias in Transformer-based Models Using Attention Guidance}

\author{Jiri Gesi}

\affiliation{%
  \institution{Amazon}
  \city{Palo Alto}
  \state{California}
  \country{USA}}
\email{jirigesi@amazon.com}

\author{Iftelhar Ahmed}
\affiliation{%
  \institution{University of California, Irvine}
  \city{Irvine}
  \state{California}
  \country{USA}}
\email{iftekha@uci.edu}

\begin{abstract}
  Transformer-based models have demonstrated considerable potential for source code modeling tasks in software engineering. However, they are limited by their dependence solely on automatic self-attention weight learning mechanisms. Previous studies have shown that these models overemphasize delimiters added by tokenizers (e.g., [CLS], [SEP]), which may lead to overlooking essential information in the original input source code. To address this challenge, we introduce SyntaGuid, a novel approach that utilizes the observation that attention weights tend to be biased towards specific source code syntax tokens and abstract syntax tree (AST) elements in fine-tuned language models when they make correct predictions. SyntaGuid facilitates the guidance of attention-weight learning, leading to improved model performance on various software engineering tasks. We evaluate the effectiveness of SyntaGuid on multiple tasks and demonstrate that it outperforms existing state-of-the-art models in overall performance without requiring additional data. Experimental result shows that SyntaGuid can improve overall performance up to 3.25\% and fix up to 28.3\% wrong predictions. Our work represents the first attempt to guide the attention of Transformer-based models towards critical source code tokens during fine-tuning, highlighting the potential for enhancing Transformer-based models in software engineering.
\end{abstract}






\maketitle

\input{01-intro.tex}

\input{02-backgrounds}
\input{03-bias_analysis}

\input{04-SyntaGuid}
\input{05-Evaluation}

\input{06-Implications}

\input{07-Related}

\input{08-threats}
\input{09-conclusion}

\bibliographystyle{ACM-Reference-Format}
\bibliography{arxiv.bib}

\end{document}

%% file: 01-intro.tex
\section{Introduction}


Pre-trained Language Models (PLMs) such as BERT~\cite{devlin2018bert}, GPT~\cite{radford2019language}, and T5~\cite{raffel2020exploring} have exhibited notable performance gains in various Natural Language Processing (NLP) tasks~\cite{clark2020electra, lample2019cross, yang2019xlnet}. This trend has been further extended to software engineering applications, including but not limited to code summarization~\cite{ahmed2022multilingual, lu2021codexglue, ahmad2021unified}, code translation~\cite{elnaggar2021codetrans, phan2021cotext, wang2021codet5}, and code search~\cite{feng2020codebert, guo2020graphcodebert, jain2020contrastive}. These models are built on the Transformer network architecture~\cite{vaswani2017attention}, featuring a self-attention mechanism that learns the weight and interdependence of attention among tokens within an input sequence.

The self-attention mechanism uses attention weight to capture inter-relationships and long-range dependencies among tokens in a sequence. Prior research has investigated how attention weights are distributed across different hidden layers of the PLM~\cite{wan2022they,feng2020codebert,guo2020graphcodebert} and different code syntax~\cite{zhang2022diet, wang2021syncobert}. In the most recent work, Zhang et al.~\cite{zhang2022diet} identified that CodeBERT~\cite{feng2020codebert} pays more attention to certain types of tokens and statements such as keywords and data-related statements. The prevailing methodology in the field involves a two-step paradigm on PLMs: pre-training followed by fine-tuning. Here, PLMs undergo fine-tuning to achieve optimal performance for specific downstream tasks. Despite this widespread usage, our understanding of how attention weight distribution evolves during the fine-tuning process and how it varies between correct and incorrect prediction subsets is limited. Prior research has illuminated that the performance of non-transformer-based models can be influenced by the biases inherent in the features they learn~\cite{gesi2021empirical}. Addressing these biases has been shown to bolster the robustness of such models~\cite{gesi2023leveraging}. In contrast, for transformer-based models, the learned features are encapsulated within these attention weights. Should biases, such as attention bias, manifest within these learned attention weights, it can pave the way for pioneering techniques that harness this knowledge, aiming to augment the efficacy of fine-tuned PLMs.


To address this research void, we investigate the disparities in attention-weight assignments between correct and incorrect sub-datasets within fine-tuned PLMs across various downstream tasks. Our empirical findings suggest that fine-tuned PLMs are inclined toward specific syntax tokens, which include identifiers, modifiers, and Abstract Syntax Tree (AST) elements like method signatures, especially during accurate predictions. Moreover, our analysis reveals a notable correlation: when PLMs allocate diminished attention to certain tokens, the performance of the fine-tuned PLMs can be adversely affected. Stemming from this observation, we introduce an innovative attention-guiding mechanism. This mechanism is designed to encourage attention heads to prioritize these pivotal syntax tokens and AST elements, thereby enhancing the efficacy of the fine-tuned model.

\begin{figure}[t]

\centering
\includegraphics[width=0.5\columnwidth]{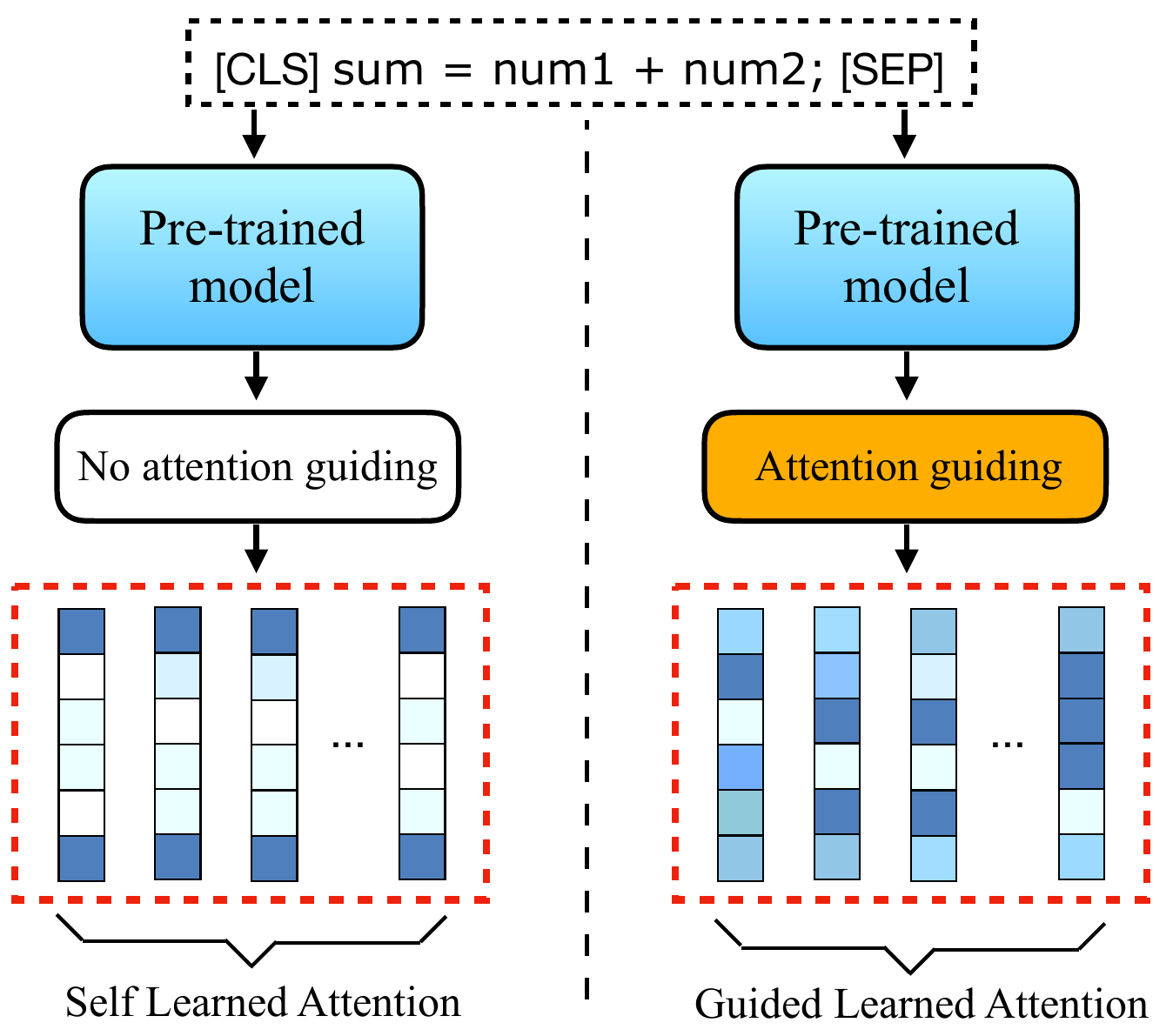}
\caption{Illustration of attention guiding mechanism}
\label{fig:ag}
\end{figure}

Figure~\ref{fig:ag} provides a visual representation of the attention-guiding mechanism we have devised for the fine-tuning of source code PLMs. On the left side of Figure~\ref{fig:ag}, one observes the input source code processed by a pre-trained model for fine-tuning, devoid of any attention-guiding interventions. The emergent attention weight vectors are autonomously derived, with darker hues signifying elevated attention weight allocations. Drawing from the insights of Sharma et al.\cite{sharma2022exploratory}, it is evident that autonomously learned attention heads frequently allocate significant attention weight to tokenization-induced delimiters, such as [CLS] and [SEP], which demarcate the initial and terminal positions in the learned attention weight vectors, respectively. In contrast, the right segment of Figure\ref{fig:ag} presents the analogous source code input, albeit with the integration of the attention-guiding mechanism. This strategy is designed to stimulate the self-attention heads to prioritize pre-designated critical tokens. A related methodology is presented by Deshpande et al.~\cite{deshpande2020guiding} for natural language text, wherein they propose generic pre-defined attention patterns to guide self-attention heads. However, due to the significant differences between programming languages and natural languages, their approach and findings are not directly applicable to software engineering tasks. To. the best of our knowledge, our work is the first to investigate attention bias and ways to mitigate it in software engineering.



In this study, we aim to investigate following research questions:

\textbf{RQ1}: How does the attention weight distribution in the fine-tuned Pre-trained language model differ for program syntax tokens and AST elements when comparing correct predictions to incorrect ones?

\textbf{RQ2}: In various downstream tasks, which attention weight allocations to syntax tokens and AST elements most profoundly influence the performance of the fine-tuned model?

\textbf{RQ3}: To what degree does our introduced Syntax pattern attention Guiding mechanism, SyntaGuid, enhance the performance of the fine-tuned Pre-trained Language Model?

Our study makes several significant contributions to the field of PLMs for software engineering:

\begin{itemize}
    \item \textbf{Attention bias investigation}: we undertook a pioneering set of experiments to provide empirical evidence highlighting the propensity of fine-tuned PLMs to exhibit bias in their attention weight towards specific source code syntax tokens and AST elements.
    \item \textbf{Attention Bias Impact Analysis}: Through rigorous experimental evaluations, we have ascertained that the performance of fine-tuned models can be adversely affected by the presence of attention bias.
    \item \textbf{Introduction of SyntaGuid}: We have introduced a cutting-edge attention-guiding mechanism, termed SyntaGuid. This technique is designed to channel the attention weight of PLMs towards pivotal source code syntax tokens and AST elements. Our empirical assessments underscore the efficacy of SyntaGuid across a spectrum of software engineering downstream tasks, cementing its stature as a versatile solution to enhance the performance of fine-tuned PLMs.
\end{itemize}

The remainder of the paper is structured as follows. Section 2 describes the necessary background. Section 3 presents details of the empirical analysis. Section 4 presents our proposed approach SyntaGuid. Section 5 places results in the broader context of work to date and Section 6 outlines the implications for practitioners and researchers. Section 7 presents the related works. Section 8 lists the threats to validate our results. Section 9 concludes with a summary of the key findings and an outlook on our future work.

%% file: 02-backgrounds.tex
\section{Background}
\label{sec:Backgrounds}

In this section, we explain the necessary backgrounds.

\subsection{Self-attention-based Transformer Model}

The Transformer~\cite{vaswani2017attention} architecture, which relies on the self-attention mechanism, has emerged as a popular choice for learning representations of source code. Let $c = \{t_1, t_2, ..., t_n\}$ denote a code snippet consisting of a sequence of $n$ code tokens. A Transformer model comprises $L$ layers of Transformer blocks that transform the code snippet into contextual representations at different layers, denoted by $H^{l} = [h_1^l, h_1^l, ..., h_n^l]$, where $l$ denotes the $l_{th}$ layer. The layer representation $H^l$ for each layer is computed using the $l_{th}$ Transformer block, i.e., $H^l = Transformer(H^{l-1}), l \in \{1, 2, ..., L\}$, where $L$ is the total number of layers. 

In each layer of the Transformer model, self-attention heads are utilized to aggregate the output vectors from the previous layer. Given an input sequence of code tokens $c = \{t_1, t_2, ..., t_n\}$, the self-attention mechanism computes a set of attention weights for each token $w_i$ over the tokens in the input, represented as:

$$Atten(w_i) = (\alpha_{i,1}(c), \alpha_{i,2}(c), ..., \alpha_{i,n}(c))$$

Here, $\alpha_{i,j}(c)$ represents the attention that token $w_i$ pays to token $w_j$, which is computed from the scaled dot-product of the query vector of $w_i$ and the key vector of $w_j$, followed by a softmax. The general form of the attention mechanism is expressed as the weighted sum of the value vector $V$, using the query vector $Q$ and the key vector $K$:

$$ Att(Q, K, V) = softmax(\frac{QK^T}{\sqrt{d_{model}}}) \cdot V$$

Here, $d_{model}$ denotes the dimensionality of the hidden representation. For self-attention, the query, key, and value vectors are obtained by mapping the previous hidden representation $H^{l-1}$ using different linear functions, i.e., $Q=H^{l-1} \cdot W_Q^{l}$, $K=H^{l-1} \cdot W_K^{l}$, and $V=H^{l-1} \cdot W_V^{l}$, respectively. Finally, the encoder produces the final contextual representation $H^{l} = [h_1^l, h_2^l, ..., h_n^l]$, which is obtained from the output of the last Transformer block.

To further clarify, the positional encoding of each token is calculated using sine and cosine functions, as shown below:

$$w_i = e(w_i) + pos(w_i)$$

where $e$ denotes the word embedding layer, and $pos$ denotes the positional embedding layer. Typically, the positional encoding implies the position of the code token based on sine and cosine functions.

Overall, the combination of self-attention mechanism, multi-head attention, and positional encoding enables Transformer models to effectively capture both the syntactic and semantic features of source code, making them a popular choice for many software engineering tasks.

\subsection{Pre-training Language Model}
Given a corpus $\mathcal{C}$, each sentence (or code snippet) is first tokenized into a series of tokens. Prior to pre-training, the model takes the concatenation of two segments as the input, defined as $c_1 = \{ t_1, t_2, ..., t_n \}$ and $c_2 = \{ w_1, w_2, ..., w_m \}$, where $n$ and $m$ denote the lengths of the two segments, respectively. The two segments are concatenated with a special separator token [SEP]. Furthermore, the first and last tokens of concatenated sequence are padded with a special classification token [CLS] and an ending token [EOS], respectively. Formally, the input of each training sample can be represented as follows:

$$ s = [CLS], t_1, t_2, ..., t_n, [SEP], w_1, w_2, ..., w_m [EOS]. $$

The Transformer encoder is then used for pre-training with two self-supervised learning objectives: Masked Language Modeling (MLM) and Next Sentence Prediction (NSP). In MLM, a certain percentage of the tokens in an input sentence is randomly selected and replaced with the special token [MASK]. Specifically, BERT chooses 15\% of the input tokens for possible replacement, and among them, 80\% are replaced with [MASK], 10\% remain unchanged, and the remaining 10\% are randomly replaced with tokens from the vocabulary. The purpose of MLM is to train the model to predict the masked tokens based on the surrounding context. For NSP, it is modeled as a binary classification task to predict whether two segments are consecutive. Positive and negative training examples are generated based on the following rules: (1) if two segments are consecutive in a document, they are considered positive examples; (2) otherwise, paired segments from different documents are considered negative examples.

\subsection{Pre-trained models for source code}

Recently, self-supervised learning techniques using MLM have gained popularity for natural language understanding and generation~\cite{gpt4, wei2021finetuned, liu2023pre}. Similarly, in the field of software engineering, several pre-trained code models have been proposed for program comprehension, code generation and etc~\cite{copilot, alphacode, codewhisperer}. In this study, we have chosen to use the CodeBERT~\cite{feng2020codebert} pre-trained model since this is one of the state-of-the-art code models for code representation learning. 

CodeBERT pre-trains the model on two tasks: MLM and Replaced Token Detection (RTD). In MLM, two random tokens from the input pair of code and natural language comments are masked, and the model aims to predict the original token from a large vocabulary. The RTD task involves two generators and a discriminator. The generators predict the original token for the masked token, while the discriminator predicts whether the tokens are original or not. After pre-training, CodeBERT can be fine-tuned on downstream tasks, making it a versatile model for a wide range of software engineering applications, such as defect detection~\cite{lu2021codexglue}, clone detection~\cite{BigCloneBentch, lu2021codexglue}, code generation~\cite{alphacode, copilot, gpt4} etc.

%% file: 03-bias_analysis.tex
\section{Empirical Analysis of Attention Weight Assignment Bias}
\label{sec:bias}

In this section, we explain the research methodology and experimental framework employed to analyze the attention bias and its impact on fine-tuned PLMs tailored for downstream tasks. The primary aim of this empirical study is to discern the degree to which fine-tuned PLMs allocate varying attention weights to specific positions within the source code syntax, contrasting between accurate and erroneous predictions. Furthermore, we seek to probe the potential ramifications arising from these disparities in attention-weight assignments. 
 
\subsection{Study Design}

Multi-head attention is a fundamental mechanism in Transformer-based models for language modeling, which enables the quantification of token importance in a given sentence. This attention distribution facilitates the learning and representation of a sentence by assigning higher weights to tokens that carry greater significance. Extracting useful information from PLMs requires an understanding of the important tokens within the code. As source code can be analyzed at different levels of granularity, including tokens, statements, and AST statements, this study focuses on the atomic unit of source code, i.e., syntax tokens and AST statements. Examining attention weights at this level of granularity can provide insights into fine-tuned PLM's performance with respect to the assigned attention on larger code blocks. Hence, we decided to focus on this granularity.

To address our research questions, we utilize the attention weights from the Transformer layers of fine-tuned PLMs to measure the importance of each token. We subsequently explore whether the collected attention weights exhibit significant differences between correct and incorrect predictions.

For instance, one of our target PLM CodeBERT consists of 12 self-attention layers, each containing 12 heads that compute attention weights for the same token. To yield a comprehensive estimate of the attention weight for each token, we adopt an approach that aggregates the attention scores across all layers and heads. This approach is consistent with prior studies in the field~\cite{sharma2022exploratory, wan2022they, ma2022self, karmakar2021pre}.


\subsection{Experiment tasks}
Our goal was to cover a wide range of tasks in our evaluation. Our analysis requires executing the PLMs for a task and analyzing the learned attention-weights differences between correct and incorrect predictions, making it difficult to analyze all possible software engineering tasks investigated in the literature. So, we elected to assess two code comprehension tasks: code clone detection and the cloze test, in addition to a code generation task termed code translation since this covers both types of tasks, namely comprehension and generation~\cite{niu2023empirical}. The subsequent sections provide an in-depth overview of each task and the corresponding dataset employed in our research. Table~\ref{tab:data} summaries of the datasets and state-of-the-art performance PLMs  for these tasks according to Niu et al~\cite{niu2023empirical}.

\begin{itemize}    

    \item \textbf{Task 1: Cloze test.} Cloze tests predict a masked word or phrase from context. Extended to source code in the CodeXGlue dataset~\cite{lu2021codexglue}, we focus on the ClozeTest-all dataset, which contains instances of masked code functions, docstrings, and 930 target words. Introduced by Microsoft Research, it covers six programming languages and is available in the CodeXGLUE repository~\cite{lu2021codexglue}.

    \item \textbf{Task 2: Clone detection.}Clone detection identifies duplicate code in software, often resulting from practices like copy-pasting, which can negatively impact software development~\cite{sajnani2016sourcerercc}. Binary classification algorithms categorize code pairs as equivalent or not. We use the BigCloneBench dataset, a comprehensive benchmark for clone detection~\cite{svajlenko2014towards}, which includes 6 million true clone pairs and 260 thousand false pairs across ten functionalities. The benchmark is available on GitHub~\cite{BigCloneBentch}.

    \item \textbf{Task3: Code translation.} This task translates software from one language to another~\cite{jiang2021cure}. We use the CodeXGLUE's Code2Code dataset, translating Java to C\#\cite{lu2021codexglue}. Curated from repositories like Lucene\cite{Lucene}, POI~\cite{POI}, and Antlr~\cite{ANTLR}, it focuses on parallel functions between languages, excluding duplicates and empty functions.
\end{itemize}

\begin{table}[t]
\centering
\caption{Details of PLMs and datasets of evaluation tasks}
\label{tab:data}
\resizebox{0.6\columnwidth}{!}{%
\begin{tabular}{ccccc}
\hline
\textbf{Task} & \textbf{\begin{tabular}[c]{@{}c@{}}Pre-trained\\ Model\end{tabular}} & \textbf{Dataset} & \textbf{Size} & \textbf{Language} \\ \hline
\begin{tabular}[c]{@{}c@{}}Cloze\\ Test\end{tabular} & CodeBERT~\cite{feng2020codebert} & CodeXGlue & 50k & Java \\ \hline
\begin{tabular}[c]{@{}c@{}}Code Clone\\ Detection\end{tabular} & SynCoBERT~\cite{wang2021syncobert} & BigCloneBench & 901k & Java \\ \hline
\begin{tabular}[c]{@{}c@{}}Code\\ Translation\end{tabular} & PLBART~\cite{ahmad2021unified} & CodeTrans & 11.5k & Java-C\# \\ \hline
\end{tabular}%
}
\end{table}

\subsection{Selected syntax types and AST statements}
For our analysis, we needed to decide on the elements for which we wanted to conduct the attention-weight analysis. We decided to analyze elements that were identified to be important to humans. Building on the foundational research by Aljehane et al.~\cite{aljehane2021determining}, which delved into the differential attention patterns between expert and novice programmers while reading source code, our investigation focuses on syntax tokens and AST elements. We focus on key syntax tokens, including \textit{identifiers, modifiers, operators, data types, separators, keywords, strings, and Booleans}. For the extraction of these tokens from the source code, we employ Javalang~\cite{javalang}, a renowned Java syntax parsing library. Additionally, we examine AST statements like \textit{Method signature, If-else, While, and Return statements} to understand the attention dynamics of the fine-tuned CodeBERT. For the identification of these AST statements, we leverage tree-sitter-java~\cite{tree-sitter}, a widely-adopted Java AST parsing library.

\subsection{Attention weight bias analysis}

In our study, we fine-tuned models for the tasks delineated earlier, employing the original training datasets and hyper-parameters as specified in the CodeXGLUE benchmark~\cite{lu2021codexglue}. Comprehensive details regarding hyper-parameters can be found on our study's companion website~\cite{AttentionBias}. Following the fine-tuning, we segregated the prediction data into two distinct categories based on the model's accuracy: those with correct predictions and those with incorrect ones. Our analysis then delved into the attention weights allocated to each syntax token and AST statement, contrasting these weights between the two aforementioned categories. It's pertinent to note that the PLM tokenizer~\cite{devlin2018bert} might fragment a single source code token into several tokens. To address this, we adopted the methodology proposed by Sharma et al~\cite{sharma2022exploratory} to compute the attention weights assigned by self-attention heads to each syntax and AST statement token. Given the multiple tests conducted, we incorporated the Bonferroni correction~\cite{bland1995multiple} to adjust for multiple hypothesis testing, leading to a revised p-value threshold of 0.01. To ascertain the presence of significant disparities between the correct and incorrect prediction groups across all syntax tokens and AST statements, we employed the non-parametric Mann-Whitney test~\cite{mann1947test}, given the non-normal distribution of our data.

\subsection{Attention bias impact analysis}

In our endeavor to discern the potential impact of attention bias on the performance of fine-tuned PLMs, we systematically partitioned the testing datasets into groups characterized by high and low attention weights. To elucidate, consider the architecture of the pre-trained CodeBERT, which comprises 12 layers, each housing 12 self-attention heads. For a given layer, we aggregated the attention weights assigned by all 12 self-attention heads across the testing instances and computed the mean value. This mean served as a threshold to categorize the attention weights of individual heads within that layer as either high or low. 

Given CodeBERT's 12-layer structure, we derived a distinct threshold for each layer. A layer was designated as 'high attention' if more than half of its self-attention heads assigned attention weights exceeding their respective thresholds. Conversely, it was labeled 'low attention' if fewer than half of the heads did so. Layers with an equal split of high and low attention heads were excluded from this categorization.

Subsequently, a testing instance was classified as receiving 'high attention' if it was processed by more than half of the layers designated as 'high attention'. Conversely, if it was processed by fewer than half of such layers, it was deemed to have received 'low attention'. Instances processed by an equal number of 'high' and 'low attention' layers were not categorized.

By employing this methodology, we effectively partitioned the testing data into paired groups of high and low attention. We then assessed the model's predictive performance across these groups to investigate the potential influence of attention bias on the efficacy of fine-tuned PLMs.

%% file: 04-SyntaGuid.tex
\section{SyntaGuid: Syntax Pattern Attention Guiding}
\label{sec:SyntaGuid}

In this section, we present our novel approach for fine-tuning Transformer-based models on source code by utilizing attention guiding. Specifically, we begin by formally defining the Masked Language Modeling (MLM) set up within the context of Transformers~\cite{vaswani2017attention} and proceed to describe the attention-guiding technique. Subsequently, we introduce our proposed Syntax Pattern Attention Guiding (SyntaGuid) technique, which leverages the syntactic structure of source code to guide the attention mechanism during fine-tuning.

\subsection{Masked Language Modeling}

The application of Transformers in sequence-to-sequence prediction tasks involves training on a dataset $\mathcal{D}$ comprising pairs of sequences $x$ and their corresponding labels $y$. In the case of MLM, the input sequence $x_1, x_2, . . . , x_n $ of length $n$ consists of individual tokens, and the output labels $y_1, y_2, . . . , y_n$ are identical to the input sequence, i.e., $y_i = x_i$. A certain fraction $k$ of the input tokens, randomly selected, are masked by replacing them with a special \textit{<MASK>} token. These masked indices are grouped together in a set $\mathcal{C}$. The MLM objective is defined as a cross-entropy loss on the model's predictions $\hat{y_i}$ at the masked locations $j \in \mathcal{C}$ and is employed to optimize all the parameters $\theta$ of the model by minimizing the loss:

\begin{equation}
    \mathcal{L}_{MLM}(x, y) = - \sum_{j \in \mathcal{c}}log\mathbb{P}(y_j|x;\theta) 
\end{equation}

The Transformer architecture used for MLM involves $l$ layers, each containing $\mathcal{h}$ self-attention heads. Let $s_k$ be the input activations to layer $k$ of this model, with $|s_k| = n$. The initial input activations $s_1$ are equivalent to the input sequence $x$, such that $s_1 = s = x$. For every position $p$ in the output, each attention head in layer $k$ induces a probability distribution over all positions in the input $s_k$. Specifically, the attention activations for a single head, denoted as a function of $s$ and described by Equation 1 in Vaswani et al.~\cite{vaswani2017attention}, can be expressed as follows:

\begin{equation}
    \mathbf{H}(s) = softmax(\frac{QK^T}{\sqrt{d_k}}) \in \mathbb{R}^{n \times n}
\end{equation}

the query and key matrices of dimension $d_k$ are denoted by $Q$ and $K$, respectively. For notational convenience, we drop the dependence on the input sequence $s$ in the following sections. The attention paid by token $p$ in the head's output layer to token $q$ in the head's input layer is represented by the scalar $\mathbf{H}(s)[p,q]$.

\begin{figure}[t]
\centering
\includegraphics[width=\textwidth]{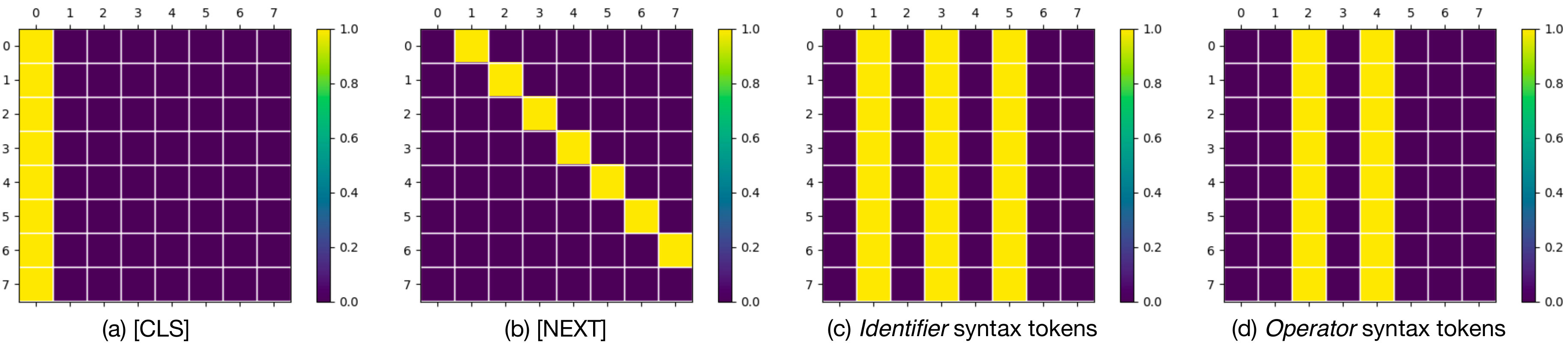}
\caption{Example attention guiding patterns for code snippet ``<s> sum = num1 + num2; <\textbackslash s>'', whose syntax type list is: [[CLS], identifier, operator, identifier, operator, identifier, separator, [SEP]].}
\label{fig:patterns}
\end{figure}

\subsection{Syntax Pattern Attention Guiding}

The technique of attention guiding has been introduced to encourage self-attention heads to allocate more attention to predefined important positions of tokens~\cite{deshpande2020guiding}. This approach can function as an auxiliary objective to regularize the fine-tuning process of downstream tasks~\cite{deshpande2020guiding, wang2022paying}. To guide an attention head, a mean squared error (MSE) loss is applied to $\mathbf{H}$ using a pre-defined pattern $\mathbf{P}(s) \equiv \mathbf{P} \in \mathbb{R}^{n \times n}$, where $|| \cdot ||_F$ denotes the Frobenius norm:

\begin{equation}
    \label{eq:ag}
    \mathcal{L}_{ag} = || \mathbf{H} - \mathbf{P} ||_F 
\end{equation}

Figure~\ref{fig:patterns} presents four examples of attention guiding patterns for a given code snippet. Specifically, Figure~\ref{fig:patterns}-(a) illustrates the attention guiding pattern that makes self-attention heads focus on the first [CLS] token. On the other hand, Figure~\ref{fig:patterns}-(b) depicts the pattern that guides self-attention heads to focus on the next tokens.

In Section~\ref{sec:bias}, our analysis of attention weights reveals a bias in the self-attention heads of fine-tuned PLMs might towards certain syntax tokens, such as identifiers and modifiers, as well as specific AST statements like method signatures.  In order to capitalize on this insight and encourage the self-attention heads to focus more on critical programming language information, we introduce two sets of syntax attention guiding patterns: syntax token attention patterns and AST statements patterns.


\begin{itemize}
    \item \textbf{Syntax token attention patterns} guide self-attention heads focusing on specific syntax type token positions, such as \textit{idetifier}, \textit{keywords}, \textit{operator}, \textit{data types} in a given source code sequence. As an example: 
    \begin{equation}
        \mathbf{P}_{Syntax}[p, q] = \left\{ \begin{array}{rcl}
                                            1 & q_{Syntax} = Identifier \\ 
                                            0 & otherwise
                                        \end{array}\right.
    \end{equation}
    where $q_{type}$ is the syntax type of source code token $q$. Two attention guiding patterns that focus on identifier and operator syntax tokens are presented in Figure~\ref{fig:patterns}-(c) and (d), respectively.

    \item \textbf{Abstract syntax tree statements attention patterns} guide attention heads focusing on token positions belong to particular AST statements, such as \textit{Method signatures}, \textit{If-else}, and \textit{Return statements}. As an example: 
    \begin{equation}
        \mathbf{P}_{AST}[p, q] = \left\{ \begin{array}{rcl}
                                            1 & q_{AST} = Return \\ 
                                            0 & otherwise
                                        \end{array}\right.
    \end{equation}
    where $q_{ast}$ belongs to the Return statements in AST for input source code sequence.  
\end{itemize}

\textbf{SyntaGuid Loss Function.} We apply the attention loss in Equation~\ref{eq:ag} to each head in each layer to obtain the overall source code syntax attention guidance (SAG) loss:

\begin{equation}
    \mathcal{L}_{SAG}(x) = \sum_{k=1}^{\ell}\sum_{j=1}^{h}\mathcal{L}_{ag} \times \mathbb{I}(k,j)
\end{equation}

where $\mathbb{I}(k,j)$ denotes an indicator function which is 1 only if the $j_th$ head in layer $k$ is being guided.

In principle, this loss permits any choice of patterns for each $\mathbf{P}{kj}$. However, for the sake of simplicity in our experiments, we guide a specific head number to the same pattern across all layers. That is, $\mathbf{P}{j}$ is constant for all layers. We use the gradients from this loss to update all the model's parameters, including the feedforward and input embedding layers. It is important to note that this loss depends solely on the input $x$ and not on the labels $y$.

Finally, the overall optimization objective is obtained by combining the attention guidance (AG) loss with the MLM loss:

\begin{equation}
    \label{eq:loss}
    \mathcal{L}(\theta) = \mathbb{E}_{(x,y) \thicksim \mathcal{D} }[\mathcal{L}_{MLM} + \alpha \cdot \mathcal{L}_{SAG}]
\end{equation}

where $\alpha$ is a hyper-parameter that controls the scale that we apply on selected heads for attention guiding. According to Deshpande et al.~\cite{deshpande2020guiding}, the $\mathcal{L}_{SAG}$ converges faster than $\mathcal{L}_{MLM}$. We linearly decay $\alpha$ from an initial value $\alpha_0 = 1$ to $0$ as the fine-tuning progresses.

\subsection{Syntax attention patterns}
Based on the attention bias results presented in Section~\ref{sec:bias}, we have observed that the self-attention heads of fine-tuned CodeBERT assign significantly higher attention weights to certain syntax tokens, including identifiers, modifiers, operators, basic data types, separators, keywords, and string tokens. Additionally, the self-attention heads of CodeBERT also assign greater attention weights to specific code structures, such as method signatures, if-else statements, and Return statements. Therefore, we propose the following attention guiding patterns for syntax token attention guiding during pre-trained model fine-tuning:

\begin{itemize}
  \item[] 1. \texttt{[Modifier]} attends to the modifier syntax tokens. 
  \item[] 2. \texttt{[Separator]} attends to the separator syntax tokens. 
  \item[] 3. \texttt{[Key]} attends to the keyword syntax tokens. 
  \item[] 4. \texttt{[Identifier]} attends to the identifier syntax tokens .
  \item[] 5. \texttt{[DataType]} attends to the basic data type syntax tokens. 
  \item[] 6. \texttt{[Operator]} attends to the operator syntax tokens. 
  \item[] 7. \texttt{[String]} attends to the string syntax tokens. 
\end{itemize}

And, following abstract syntax tree attention guiding patterns:

\begin{itemize}
  \item[] 1. \texttt{[MethodSignature]} attends to tokens belonging to the method signature AST statement. 
  \item[] 2. \texttt{[IfElseElement]} attends to tokens belonging to the If-else AST statement. 
  \item[] 3. \texttt{[ReturnElement]} attends to tokens belonging to the Return AST statement. 
\end{itemize}

Furthermore, to enable a comprehensive evaluation of our proposed attention guiding patterns, we compare their performance with the local and global attention patterns proposed by Deshpande et al.~\cite{deshpande2020guiding}. The global attention patterns focus self-attention heads on global position, such as \texttt{[First]}, \texttt{[CLS]}, and \texttt{[SEP]}. And the local attention patterns either focus on the next or previous tokens, such as [NEXT] and [PREV]. This enables us to conduct a fair and thorough analysis of the effectiveness of our proposed attention guiding patterns compared to other existing patterns.

\subsection{Experimental setup}

In this study, we employ the software engineering tasks of code clone detection, cloze test, and code translation, which were previously described in Section~\ref{sec:bias}, along with their respective datasets. It is noteworthy that the AG patterns proposed by Deshpande et al.~\cite{deshpande2020guiding} are intended for natural languages, whereas our guiding patterns are specifically designed for source code. As a result, we replicated their AG patterns to compare their effectiveness with our code-specific patterns.

\subsubsection{Implementation details}

To ensure comparability across different experimental settings, we select CodeBERT~\cite{feng2020codebert} as the foundational pre-trained model for all our evaluations. CodeBERT is a prominent pre-trained model specifically designed for code, and has served as the foundation for RoBERTa~\cite{liu2019roberta}, a widely-used language model in other programming language modeling studies~\cite{guo2020graphcodebert, kanade2020learning, xu2022systematic, wang2022bridging, wan2022they, wang2021syncobert}.

\textbf{Basic model fine-tune.} We tune the learning rate is 5e-5 for two epochs. The batch size for training is 16 and for testing is 32.

\textbf{AG model.} For attention guiding models, we guide a fraction of $\lambda \in \{\frac{1}{4}, \frac{2}{4}, \frac{3}{4}, 1 \}$ of heads in each layer. We choose $\alpha$ for equation~\ref{eq:loss} from the set $\{ 1, 10, 100 \}$ such that scales of the MLM loss and auxiliary loss are comparable at the beginning of the fine-tuning. To achieve fair comparison and reduce deep learning model's variance impact~\cite{pham2021deviate}, we used five-fold cross validation for each basic and AG model. 

\subsubsection{Evaluation metrics}

Consistent with prior works on code clone detection~\cite{jiang2007deckard, white2016deep, wang2020detecting, feng2020codebert}, we evaluate the performance of our models using precision, recall, and F1 score~\cite{goutte2005probabilistic}. Precision measures the accuracy of the predicted clone pairs. Whereas recall represents the proportion of actual clone pairs correctly predicted by the model. The F1 score is the harmonic mean of precision and recall, providing a balanced assessment of the model's performance.

The objective of the Cloze test is to predict the appropriate code token for a blank position in the context of the surrounding code. Consequently, we evaluate the prediction accuracy, which is calculated using the same formula as precision (i.e., the number of correct predictions divided by the total number of predictions).

Regarding the code translation task, we adopt the evaluation metrics proposed in CodeXGLUE. Specifically, we report three metrics: BLEU~\cite{papineni2002bleu} score, CodeBLEU score~\cite{lu2021codexglue}, and accuracy (ACC). BLEU score is a commonly used metric for machine translation tasks, which measures the similarity between the generated code and the target code based on n-gram precision. CodeBLEU is a variant of BLEU proposed by CodeXGLUE, which takes into account not only surface-level matching but also grammatical and logical correctness, utilizing the AST and data-flow structure. In addition, we also evaluate the accuracy, which calculates the exact match between generated code and the target code.

%% file: 05-Evaluation.tex
\section{Results}
\label{sec:evaluation}

In this section, we delineate the results of our experiments pertaining to the subsequent research questions:

\begin{itemize}
\item \textbf{RQ1}: How does the attention weight distribution in the fine-tuned PLMs differ for program syntax tokens and AST statements when comparing correct predictions to incorrect ones?
\item \textbf{RQ2}: In various downstream tasks, which attention weight allocations to syntax tokens and AST statements most profoundly influence the performance of the fine-tuned model?
\item \textbf{RQ3}: To what degree does our introduced syntax pattern attention guiding mechanism, SyntaGuid, enhance the performance of the fine-tuned Pre-trained Language Model?
\end{itemize}

Due to space constraints, we present only the results for CodeBERT on the cloze test across all research questions in this paper. The evaluation results for code clone detection and code translation tasks are available on our companion website~\cite{AttentionBias} which shows similar results as the cloze test.

\subsection{Attention Bias in Fine-Tuned PLMs}

\begin{figure*}[t]
    \centering
    \includegraphics[width=\textwidth]{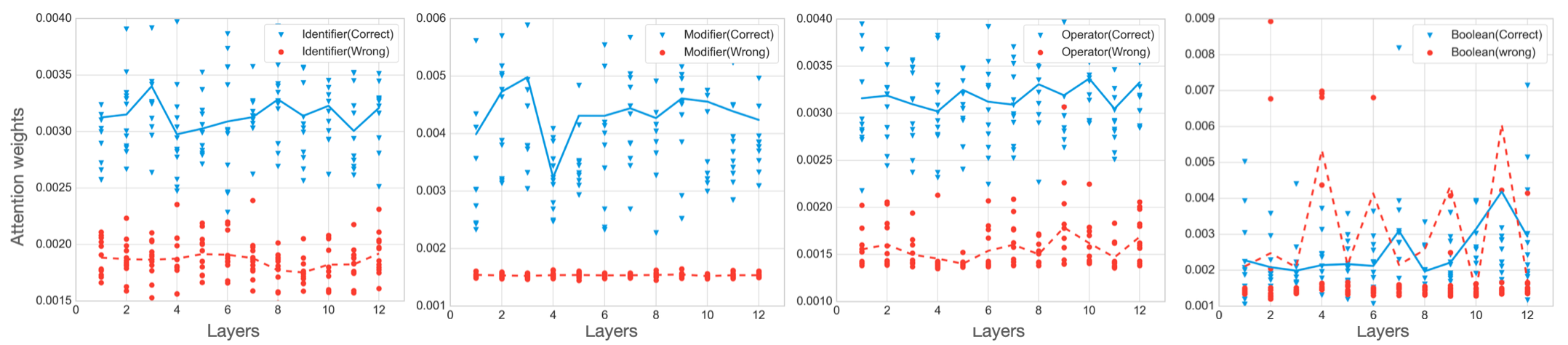}
    \caption{Comparison of attention weights on syntax tokens: Correctly Predicted vs. Mis-predicted groups}
    \label{fig:syntax_bias}
\end{figure*}

\begin{figure*}[t]
    \centering
    \includegraphics[width=\textwidth]{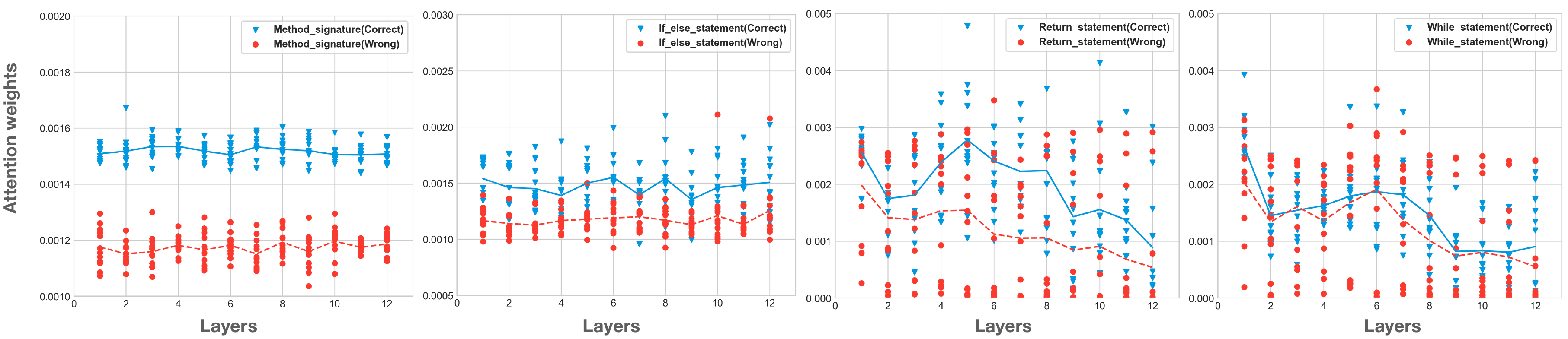}
    \caption{Comparison of attention weights on AST statements: Correctly Predicted vs. Mis-predicted groups}
    \label{fig:ast_bias}
\end{figure*}

Our primary objective was to ascertain whether attention bias exists in fine-tuned PLMs for software engineering tasks. We partitioned the test data into subsets of correctly predicted and mispredicted instances and then assessed potential disparities in attention assignment to source code syntax and AST element tokens. This analysis was crucial in determining the presence or absence of attention bias in these models.

Figure~\ref{fig:syntax_bias} and \ref{fig:ast_bias} depict the attention weights allocated by the 12 attention heads across 12 layers to various syntax tokens and AST statements for cloze test\cite{lu2021codexglue}. In Figure~\ref{fig:syntax_bias} and \ref{fig:ast_bias}, the triangle \ding{116} symbolizes the average attention values designated to syntax tokens for correctly predicted instances, with the solid line representing the cumulative average of all attention heads. Conversely, the circle \ding{108} represents the average attention values for mispredicted instances, with the dashed line indicating the overall average across all heads.

The data in Figure~\ref{fig:syntax_bias} suggests that, for successful predictions in the cloze test, the self-attention mechanism tends to favor certain syntax tokens, including Identifier, Modifier, Operators and etc. However, attention weights for Boolean syntax tokens do not exhibit a pronounced difference. To ascertain the statistical significance of these disparities, we conducted a paired t-test~\cite{mann1947test}, comparing attention weights between correctly and incorrectly predicted instances. The results indicate statistically significant differences in attention weights for Identifier (p-value < 2.13e-20), Modifier (p-value < 1.12e-16) and Operator (p-value < 4.82e-21) syntax tokens. In contrast, Boolean tokens did not exhibit a significant difference (p-value < 0.31). The self-attention mechanism within the fine-tuned PLM distinctly exhibits a propensity to allocate heightened attention weights to syntax tokens, notably the Identifier, Modifier, and Operator.

As depicted in Figure~\ref{fig:ast_bias}, the self-attention heads manifest a pronounced inclination to allocate higher attention weights to specific AST statements, notably Method Signatures, If-else, and Return statements, when predictions by the fine-tuned PLM model are accurate. A statistical evaluation was undertaken to discern the disparities in attention distribution across these AST statements between correct and erroneous predictions. The analysis revealed significant differences in attention for Method Signatures (p-value < 4.70e-26), If-else statements (p-value < 2.43e-12), and Return statements (p-value < 0.92e-3). Conversely, attention distribution for While statement did not exhibit a significant difference (p-value < 0.36). The fine-tuned PLM's self-attention mechanism notably favors AST statements such as Method Signatures, If-else and Return statements.

\subsection{Evaluation of Model Performance Under Attention Bias}

\begin{figure}[t]
    \centering
    \includegraphics[width=0.85\columnwidth]{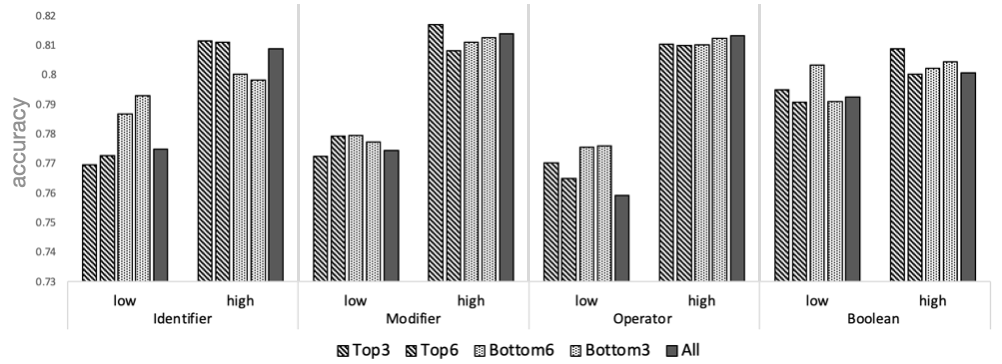}
    \caption{Comparison of model accuracy based on Syntax attention weight: Low vs. High attention weights}
    \label{fig:rq2}
\end{figure}

In addressing our second research question, we investigated the influence of attention bias on the performance of fine-tuned PLMs. To this end, we partitioned the test dataset into subsets characterized by high and low attention weights assigned to source code syntax and AST tokens using the approach explained in Section 3.5. The ensuing performance disparities between these subsets were then evaluated for each program syntax token.

Figure~\ref{fig:rq2} offers a comparative analysis of model accuracy across different syntax tokens based on the attention weights assigned. The dataset was categorized based on a spectrum of attention weight thresholds, derived from the attention weights of all 12 heads in each layer. This figure particularly underscores the model's performance on four distinct syntax tokens. From Figure~\ref{fig:rq2}, it is evident that the model tends to perform better when it allocates more pronounced attention weights to specific syntax tokens, notably the Identifier, Modifier, and Operator tokens, with accuracy drops being most pronounced when all 12 heads allocate diminished attention. However, this trend is less pronounced for Boolean tokens, where the performance differential is relatively muted.

\begin{figure}[t]
    \centering
    \includegraphics[width=0.85\columnwidth]{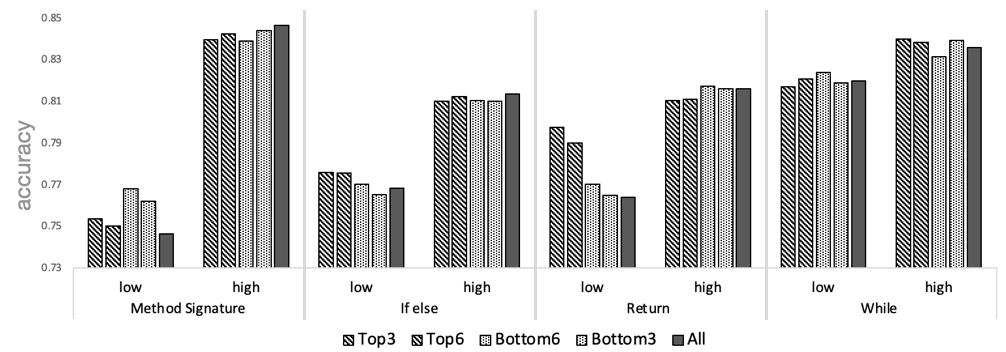}
    \caption{Comparison of model accuracy based on AST attention weight: Low vs. High attention weights}
    \label{fig:rq2-2}
\end{figure}

Figure~\ref{fig:rq2-2} offers a comparative analysis of model accuracy based on the attention weights allocated to various AST statements. The model's performance drops most significantly when it allocates diminished attention to the "Method Signature" statement, with the accuracy dropping from 84.62\% to 74.83\%. For "If-else" and "Return" statements, the most pronounced accuracy drops are observed when all 12 heads allocate diminished attention, resulting in drops from 81.32\% to 76.38\% and 81.58\% to 76.38\%, respectively. The "While" statement, on the other hand, exhibits the least change in accuracy, with the most significant drop being from 83.95\% to 81.68\% when the top 3 heads allocate diminished attention.

Our findings highlight that the performance of fine-tuned PLMs can be significantly influenced by the attention dynamics directed towards specific source code syntax tokens and AST statements, with the ``Method Signature'' emerging as a particularly influential element.

\subsection{Evaluation of Syntax Pattern Attention Guiding}

Our previous findings indicate that attention bias does influence fine-tuned PLMs (Ft-PLMs) performance when self-attention heads assign lower attention weight to particular syntax and AST tokens, leading to our proposal of the SyntaGuid attention guiding (AG) technique, which can push part of self-attention heads pay extra attention on particular tokens. To ascertain the efficacy of SyntaGuid in mitigating performance issues in attention-biased subsets, we conducted further experiments, the results of which are presented herein. 

Table~\ref{tab:rq3-1} presents the empirical results on three software engineering tasks. The baseline fine-tuned PLMs on three tasks without using any attention guiding techniques. The syntax and AST attention guiding patterns are proposed in this study and the global and local AG patterns are derived from Deshpande et al.~\cite{deshpande2020guiding}.

In cloze test, our experimental results reveal that AG patterns enhance Ft-PLM's predictive accuracy. Specifically, the application of global AG patterns improves Ft-PLM's prediction accuracy from 64.57\% to 64.71\%, while local attention patterns lead to an accuracy improvement to 64.86\%. Notably, when both local and global AG patterns are utilized concurrently, the resulting accuracy is further enhanced to 64.95\%. In contrast, the incorporation of syntax token AG patterns results in a significant improvement in Ft-PLM's prediction accuracy, achieving a score of 65.88\% (p-value < 5.24e-07). Similarly, the utilization of AST AG patterns leads to an accuracy improvement of 66.27\% (p-value < 1.94e-08). Moreover, the simultaneous integration of syntax token and AST AG patterns results in an accuracy improvement to 67.82\% (p-value < 3.19e-10).

\begin{table}[]
\centering
\caption{Results on software engineering tasks. 'AG' denotes attention guiding patterns; $\mathbf{AG_{global}}$ and $\mathbf{AG_{local}}$ are from~\cite{deshpande2020guiding}, while $\mathbf{AG_{syntax}}$ and $\mathbf{AG_{AST}}$ are introduced in this study. Numbers with * are statistically significant (paired t-test) compared to their respective Ft-PLM values. }
\label{tab:rq3-1}
\resizebox{\textwidth}{!}{%
\begin{tabular}{cccccccccccccc}
\hline
\multicolumn{1}{c}{\textbf{Task name}} &  & \multicolumn{2}{c}{\textbf{\begin{tabular}[c]{@{}c@{}}Cloze\\ test\end{tabular}}} & \textbf{} & \multicolumn{4}{c}{\textbf{\begin{tabular}[c]{@{}c@{}}Code clone\\ detetcion\end{tabular}}} & \textbf{} & \multicolumn{4}{c}{\textbf{\begin{tabular}[c]{@{}c@{}}Code\\translation\end{tabular}}} \\ \cline{1-1} \cline{3-4} \cline{6-9} \cline{11-14} 
\multicolumn{1}{c}{\textbf{\begin{tabular}[c]{@{}c@{}}Evaluation\\ metrics\end{tabular}}} &  & \textbf{Acc.} & \textbf{\begin{tabular}[c]{@{}c@{}}Acc. Delta\\ (\%)\end{tabular}} & \textbf{} & \textbf{Pre.} & \textbf{Rec.} & \textbf{F1} & \textbf{\begin{tabular}[c]{@{}c@{}}F1 Delta\\ (\%)\end{tabular}} & \textbf{} & \textbf{BLEU} & \multicolumn{1}{l}{\textbf{CodeBLEU}} & \textbf{Acc.} & \textbf{\begin{tabular}[c]{@{}c@{}}Acc. delta\\ (\%)\end{tabular}} \\ \cline{1-1} \cline{3-4} \cline{6-9} \cline{11-14} 
Ft-PLMs &  & 64.57 & - &  & 0.947 & 0.935 & 0.941 & - &  & 71.99 & 85.10 & 59.00 & - \\ \cline{1-1} \cline{3-4} \cline{6-9} \cline{11-14} 
Ft-PLMs + $AG_{global}$ &  & 64.71 & 0.14 &  & 0.951 & 0.933 & 0.942 & 0.10 &  & 74.83 & 85.35 & 59.31 & 0.3 \\
Ft-PLMs + $AG_{local}$ &  & 64.86 & 0.29 &  & 0.935 & 0.940 & 0.938 & -0.34 &  & 72.05 & 86.79 & 59.73 & 0.7 \\
Ft-PLMs + $AG_{global}$ + $AG_{local}$ &  & 64.95 & 0.38 &  & 0.948 & 0.947 & 0.948 & 0.70 &  & 73.95 & 86.73 & 60.21 & 1.21 \\ \cline{1-1} \cline{3-4} \cline{6-9} \cline{11-14} 
Ft-PLMs + $AG_{syntax}$ &  & \textbf{65.88*} & 1.31 &  & 0.959 & 0.934 & 0.946 & 0.54 &  & 74.36 & 87.82 & 60.55 & 1.55 \\
Ft-PLMs + $AG_{AST}$ &  & \textbf{66.27*} & 1.70 & \multicolumn{1}{l}{\textbf{}} & 0.954 & 0.931 & 0.942 & 0.13 &  & 72.91  & 86.55 & 60.82 & 1.8 \\
Ft-PLMs + $AG_{syntax}$ + $AG_{AST}$ &  & \textbf{67.82*} & 3.25 &  & \textbf{0.962*} & 0.938 & \textbf{0.950*} & 0.88 &  & \textbf{76.88*} & \textbf{88.23*} & \textbf{61.93*} & 2.9 \\ \hline
\end{tabular}%
}
\end{table}

In our investigation of code clone detection, we found that the use of global AG patterns improved the F1 score of Ft-PLM from 0.941 to 0.942, while the application of local attention patterns resulted in a decrease in F1 score to 0.938. However, when both global and local attention patterns were applied simultaneously, the F1 score increased to 0.948. Interestingly, we observed that the use of our proposed syntax token and AST statements attention patterns resulted in a higher F1 score of 0.946 and 0.942, respectively. When both sets of attention patterns were applied simultaneously, the F1 score further increased to 0.950 (p-value < 6.53e-05). It is worth mentioning that the default Ft-PLM already achieved a very high F1 score of 0.941 in code clone detection, and the addition of AG patterns only resulted in a marginal improvement.

In the task of code translation, our empirical results show that the application of global AG patterns enhances the BLEU score of the default Ft-PLM model from 71.99 to 74.83, CodeBLEU score from 85.10 to 85.35 and accuracy from 59.00 to 59.31. Similarly, local AG patterns improve the BLEU score to 72.05, CodeBLEU score to 86.79, and accuracy to 59.73. By using both patterns simultaneously, we achieve a further improvement in BLEU score (73.95), CodeBLEU score (86.73) and accuracy (60.21). In contrast, our proposed syntax token AG pattern improves the default Ft-PLM's code to code translation BLEU score to 74.36, CodeBLEU to 87.82, and accuracy to 60.55. Furthermore, the AST AG pattern enhances the BLEU score to 72.91, CodeBLEU to 86.55, and accuracy to 60.82. Finally, when both patterns are applied to the Ft-PLM model, the BLEU score reaches 76.88, CodeBLEU score reaches 88.23 and accuracy improves to 61.93.

\begin{table*}[]
\centering
\caption{Attention guiding performance on fixing wrong predictions by default Ft-PLMs}
\label{tab:rq11}
\resizebox{\textwidth}{!}{%
\begin{tabular}{cccccclccccccccc}
\hline
\multirow{2}{*}{\textbf{\begin{tabular}[c]{@{}c@{}}Task\\ Name\end{tabular}}} &  & \multicolumn{4}{c}{\textbf{\begin{tabular}[c]{@{}c@{}}Cloze\\ test\end{tabular}}} & \textbf{} & \multicolumn{4}{c}{\textbf{\begin{tabular}[c]{@{}c@{}}Code clone \\ detection\end{tabular}}} & \multicolumn{1}{c}{\textbf{}} & \multicolumn{4}{c}{\textbf{\begin{tabular}[c]{@{}c@{}}Code\\ Translation\end{tabular}}} \\ \cline{3-6} \cline{8-11} \cline{13-16} 
 &  & \begin{tabular}[c]{@{}c@{}}Correct\\ prediction\end{tabular} & \begin{tabular}[c]{@{}c@{}}Wrong\\ prediction\end{tabular} & \begin{tabular}[c]{@{}c@{}}Fixed \\ prediction\end{tabular} & \begin{tabular}[c]{@{}c@{}}Fix \\ \% \end{tabular} &  & \begin{tabular}[c]{@{}c@{}}Correct\\ prediction\end{tabular} & \begin{tabular}[c]{@{}c@{}}Wrong\\ prediction\end{tabular} & \begin{tabular}[c]{@{}c@{}}Fixed \\ prediction\end{tabular} & \begin{tabular}[c]{@{}c@{}}Fix\\ \% \end{tabular} &  & \begin{tabular}[c]{@{}c@{}}Correct\\ prediction\end{tabular} & \begin{tabular}[c]{@{}c@{}}Wrong\\ prediction\end{tabular} & \begin{tabular}[c]{@{}c@{}}Fixed \\ prediction \end{tabular} & \begin{tabular}[c]{@{}c@{}}Fix \\ \% \end{tabular} \\ \cline{1-1} \cline{3-6} \cline{8-11} \cline{13-16} 
Ft-PLMs &  & 2,412 & 1,323 & - &  &  & 393,399 & 22,017 & - & - &  & 590 & 410 & - &  \\ \cline{1-1} \cline{3-6} \cline{8-11} \cline{13-16} 
Ft-PLMs + $AG_{global}$ &  & 2,417 & 1,318 & 5 & 0.40\% &  & 395,061 & 20,355 & 1,662 & 7.55\% &  & 593 & 407 & -3 & 0.76\% \\ \cline{1-1} \cline{3-6} \cline{8-11} \cline{13-16} 
Ft-PLMs + $AG_{local}$ &  & 2,423 & 1,312 & 11 & 0.82\% &  & 388,414 & 27,002 & -4,985 & -22.64\% &  & 597 & 403 & -7 & 1.78\% \\ \cline{1-1} \cline{3-6} \cline{8-11} \cline{13-16} 
\begin{tabular}[c]{@{}c@{}}Ft-PLMs + $AG_{global}$ \\ + $AG_{local}$ \end{tabular} &  & 2,426 & 1,309 & 14 & 1.07\% &  & 392,568 & 22,848 & -831 & -3.77\% &  & 602 & 398 & -12 & 2.95\% \\ \cline{1-1} \cline{3-6} \cline{8-11} \cline{13-16} 
Ft-PLMs + $AG_{syntax}$ &  & 2,461 & 1,274 & 49 & 3.70\% &  & 398,259 & 17,157 & 4,860 & 22.08\% &  & 606 & 395 & -16 & 3.78\% \\ \cline{1-1} \cline{3-6} \cline{8-11} \cline{13-16} 
Ft-PLMs + $AG_{AST}$ &  & 2,475 & 1,260 & 63 & 4.80\% &  & 396,431 & 18,985 & 3,033 & 13.77\% &  & 608 & 392 & -18 & 4.44\% \\ \cline{1-1} \cline{3-6} \cline{8-11} \cline{13-16} 
\begin{tabular}[c]{@{}c@{}}Ft-PLMs + $AG_{syntax}$\\ + $AG_{AST}$ \end{tabular} &  & 2,533 & 1,202 & 121 & \textbf{9.17\%} &  & 399,630 & 15,786 & 6,231 & \textbf{28.30\%} &  & 619 & 381 & -29 & \textbf{7.15\%} \\ \hline
\end{tabular}%
}
\end{table*}

Our proposed attention guiding mechanism aims to improve Ft-PLM prediction performance by fixing incorrect predictions by the model. In addition to evaluating the performance of the fine-tuned models using commonly used evaluation metrics for each software engineering task, we also investigated the effectiveness of the AG patterns in rectifying mispredicted samples. Table~\ref{tab:rq11} presents an overview of the incorrectly predicted instances by the default Ft-PLM model and the fine-tuned models incorporating various AG patterns. Notably, for the cloze test, our proposed syntax token AG and AST AG patterns were successful in rectifying 9.17\% of the mispredicted instances, outperforming the 1.07\% instances corrected by the global and local AG patterns. For code clone detection, our proposed syntax attention patterns were effective in fixing 28.3\% of the mispredicted instances. Interestingly, the global and local attention patterns resulted in an increased 3.77\% of the mispredicted instances. Finally, for code translation, our proposed syntax AG patterns led to the correction of 7.15\% of the mispredicted instances, whereas the global and local attention patterns only corrected 3.78\% of the mispredicted instances. Syntax token and AST attention guiding patterns have demonstrated significant performance improvements over the default Ft-PLM model. Notably, these patterns have exhibited better performance on software engineering tasks than previously proposed global and local attention guiding patterns.

\begin{figure}[t]
\centering
\includegraphics[width=0.65\columnwidth]{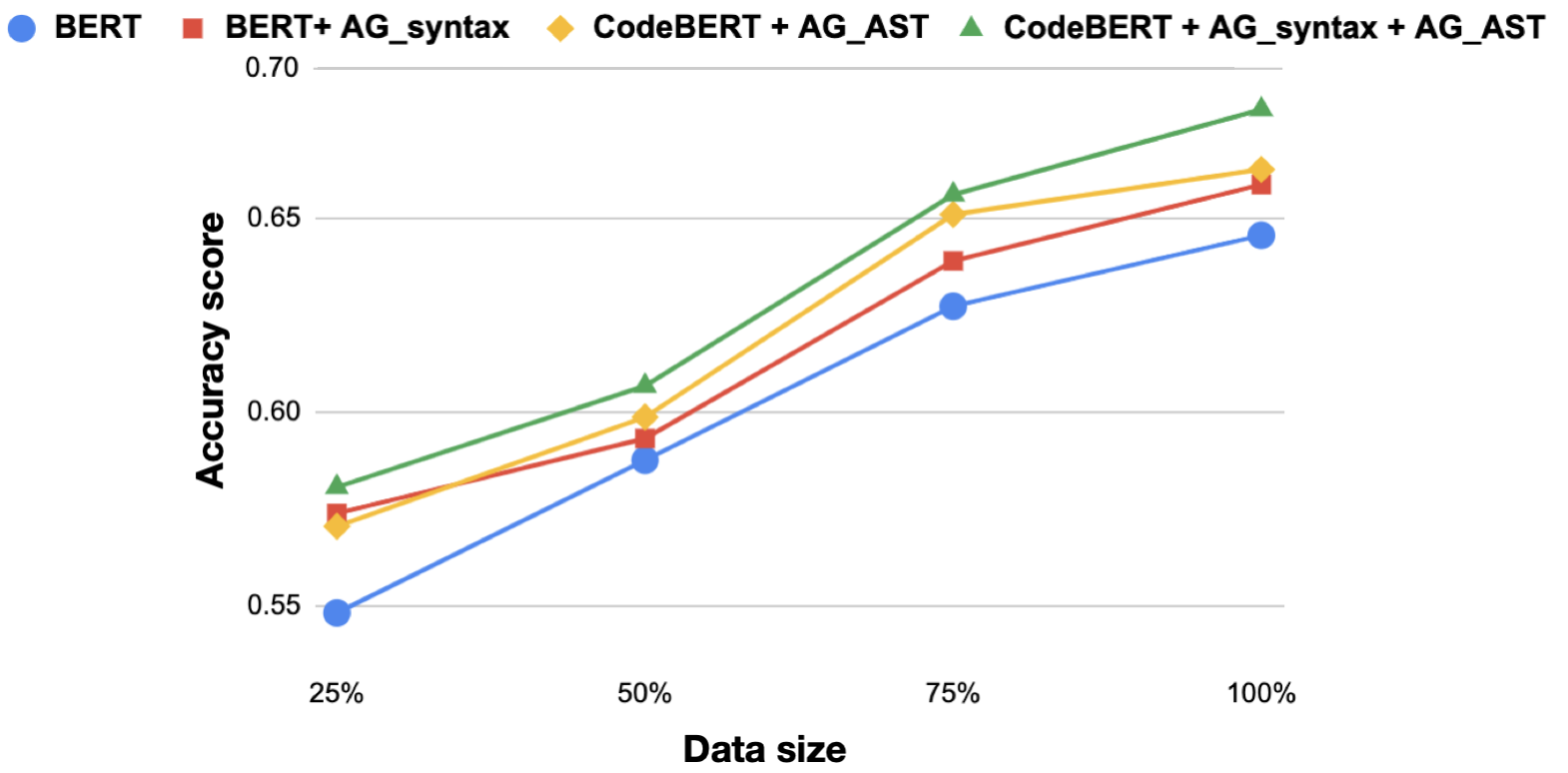}
\caption{Performance of Ft-PLM with syntax AG patterns across training data sizes}
\label{fig:size}
\end{figure}

One interesting observation from Table~\ref{tab:rq3-1} was that the efficacy of the proposed syntax token and AST statements AG patterns is more prominent for the Cloze test as compared to code translation. We posit that this is due to the relatively larger data size of Cloze test (50k) as compared to code translation (around 11.5k). So we sought to investigate the impact of training data size on the proposed AG patterns' effectiveness. For this purpose, we randomly selected 25\%, 50\%, 75\%, and 100\% of the training set for Cloze test fine-tuning and conducted experiments. Our findings indicate that both syntax token AG patterns and AST AG patterns improve the fine-tuned PLM model's performance on the Cloze test at different training data sizes. Combining both AG patterns always resulted in the best performance. When only 25\% of training data was used, syntax token AG patterns performed better than AST AG patterns. However, after using 50\% of training data, the AST AG patterns consistently outperformed the syntax token AG patterns. For a detailed breakdown of our findings, we refer the reader to Figure~\ref{fig:size}.

Based on our experimental results, we can conclude that our proposed SyntaGuid is able to improve the fine tuned PLMs performance through pushing part of self-attention heads paying more attention to part of syntax and AST statement tokens at different training data size.

%% file: 06-Implications.tex
\section{Implications}
\label{sec:discussion}
Based on our findings and analyses, we provide the following implications for researchers and practitioners.
\subsection{Implications for researchers}

We demonstrate that fine-tuned PLM assigns significantly greater weights to specific types of syntax tokens and AST elements when making correct predictions in Section~\ref{sec:bias}. This provides a new perspective for interpreting attention-based models and analyzing the attention weight distribution in Transformer-based models. One intriguing future research entails investigating the applicability of these syntax tokens and AST elements and their associated weights for building defect prediction models. Another interesting future research direction would entail utilizing this information to build a tool for explaining the model’s decisions to a developer. Also, we believe a study on a more extensive group of software engineering tasks and language models can uncover more syntax tokens and AST elements and opportunities to further improve the performances of fine-tuned models. 

We propose a method to guide self-attention heads to pay more attention to critical token positions, though our approach only guides a fraction of the self-attention heads to focus on input source code sequence positions. However, there is potential for more fine-grained attention guidance, such as directing identifier tokens to pay attention to operator tokens or if-else elements. Such granular token-to-token attention analysis may prove valuable for guiding attention. Additionally, researchers have proposed utilizing contrastive learning methods or pruning unimportant source code to retrain the pre-trained language models. Combining syntax-based attention guidance with other methods may lead to further improvements in Transformer-based models for programming languages.

\subsection{Implications for practitioners}
Based on the results presented in Table~\ref{tab:rq11}, it is evident that incorporating syntax attention guiding patterns can effectively rectify erroneous predictions without adding extra data. Oversampling has been demonstrated to be an effective approach for enhancing the performance of machine learning models. However, utilizing Transformer-based models to automatically learn representation features for programming languages presents a challenge in achieving oversampling, as the learned representation features are challenging to employ for this purpose, in contrast to tabular data. Hence, our proposed attention guiding mechanism, which does not require any extra data, represents a promising option for improving Transformer-based models for software engineering tasks. Moreover, as illustrated in Figure~\ref{fig:size}, our attention guiding approach exhibits a robust performance across different data sizes, further highlighting its efficacy.

%% file: 07-Related.tex
\section{Related Works}
\label{sec:related}

\subsection{Attention analysis in Software Engineering}
Attention-based neural networks have found extensive applications in the realm of software engineering~\cite{leclair2019neural,alon2018code2seq, haque2020improved,li2019improving,shuai2020improving}. This attention mechanism plays a crucial role by assigning heightened attention values or energy to salient elements within input data, enabling researchers to construct models or provide insights into their predictions. This mechanism has been harnessed in diverse ways within the field, such as for code summarization~\cite{leclair2019neural}, for the selection of vital Abstract Syntax Tree paths to predict appropriate function names~\cite{alon2018code2seq}, for identifying pertinent sections in context files to generate summaries ~\cite{haque2020improved}, in the context of code search~\cite{shuai2020improving}, and for enhancing bug detection~\cite{li2019improving}. Our approach distinguishes itself from previous works in a notable manner. Rather than solely utilizing the attention mechanism for training a neural model, we embark on a comprehensive exploration and analysis of the attention mechanism itself. This exploration aims to uncover any inherent biases, shed light on its behavior and characteristics, and subsequently propose and validate an approach to mitigate these biases. Our overarching goal is to enhance the overall performance of the model by addressing these attention-related biases.

\subsection{Analyzing self-attention weight}

Recent studies~\cite{ahmad2021unified,mastropaolo2021studying, paltenghi2021thinking, karmakar2021pre} have investigated the attention assignment patterns of Transformer-based language models trained for software engineering tasks. For instance, Karmakar etal~\cite{karmakar2021pre} applied four probing tasks on pre-trained code models to investigate whether pre-trained models can learn different aspects of source code such as syntactic, structural, surface-level, and semantic information. Wan et al.~\cite{wan2022they} showed that CodeBERT's attention aligns strongly with syntax structure of the code and it preserves the syntax structure of code in the intermediate representations of each Transformer layer. In addition, Zhang et al~\cite{zhang2022diet} and Sharma et al~\cite{sharma2022exploratory} reveal that CodeBERT, in general, pays more attention to certain types of tokens and statements. However, none of the prior studies investigated the alteration of attention weight distribution
between correct and incorrect prediction groups and our study sets itself apart by showing that CodeBERT demonstrates a noteworthy bias toward assigning greater attention weights to particular syntax tokens and statements when making correct predictions. 

\subsection{Guiding self-attention weight}
Numerous studies have delved into techniques for directing the self-attention mechanisms of language models toward crucial syntax tokens and statements within source code. For example, Zhang et al. \cite{zhang2022diet} introduced a novel pre-trained model for source code that steers attention towards vital syntax tokens by excluding less important or frequently occurring tokens in the input source code sequence during the pre-training phase. Similarly, Wang et al. \cite{wang2021syncobert} proposed an alternative pre-trained model that channels the attention of self-attention heads towards symbolic and syntactic characteristics of source code through contrastive learning \cite{khosla2020supervised}. In contrast to these approaches, which primarily focus on the pre-training phase, our method centers on attention guidance during the fine-tuning stage. This approach can enhance the performance of the fine-tuned model more efficiently, as fine-tuning demands considerably less time, computational resources, and data compared to the resource-intensive pre-training phase.


%% file: 08-threats.tex
\section{Threats to Validity}
\label{sec:threats}

We have taken care to ensure that our results are unbiased and have tried to eliminate the effects of random noise, but it's possible that our mitigation strategies may not have been effective. 

\textbf{Dataset Bias:}
It is important to note that our findings may not necessarily apply to all software engineering datasets and tasks, as we have only evaluated our approach on the publicly available BigCloneBench~\cite{wang2020detecting}, CodeXGLUE code translation, and cloze test datasets~\cite{lu2021codexglue}. However, these datasets have been used in previous studies~\cite{le2022coderl, yang2022natural, wang2022bridging, chen2022varclr, zhang2022diet}. Thus, their quality and reliability are well-established. We are aware that concerns have been raised regarding BigCloneBench; hence we conducted the studies on other datasets to mitigate any bias arising due to conducting the study only on BigCloneBench dataset. Moreover, the software engineering datasets we have considered are diverse in size, programming language, and complexity, which mitigates concerns of bias due to dataset selection. Therefore, we believe our selection is appropriate to address the research questions.

\textbf{Bias Due to Syntax Extraction:} 
One potential bias can arise from the syntax extraction process. Specifically, we utilized javalang~\cite{javalang} to extract the source code token syntax types and tree-sitter-java~\cite{tree-sitter} to extract the AST elements, and the selected sets of syntax types and structures were obtained from Aljehane et al.~\cite{aljehane2021determining} study about the attention difference between expert and novice programmers' when debugging. Even though different syntax extraction libraries may yield different results, the libraries we used have been widely adopted in previous research~\cite{sharma2022exploratory, le2022hyperast, svyatkovskiy2020intellicode}.

\textbf{Bias Due to Pre-trained Language Model:}
Our study focuses on examining the attention-weight assignment differences in fine-tuned language models. Despite the existence of various PLMs, including GraphCodeBERT, SynCoBERT, and CodeT5, we chose to employ CodeBERT. This decision was influenced by the fact that prior studies referenced in the related work section (Section~\ref{sec:related}) conducted their analyses using CodeBERT. We aimed to facilitate direct comparisons and contrasts between our findings and theirs.

\textbf{Bias Due to Implementation:} 
To address the potential bias, we took several measures to minimize errors in our study. First, we relied on existing implementations in CodeXGLUE for fine-tuning CodeBERT without attention guiding and applying global and local attention guiding patterns, as well as our proposed syntax attention guiding patterns. Additionally, we thoroughly tested our code and data to identify and correct any potential errors. However, we cannot completely rule out the possibility of implementation bias.

%% file: 09-conclusion.tex
\section{Conclusion}
\label{sec:conclusion}

In our research, we embarked on an exploration to discern the allocation patterns of attention weights by self-attention heads in Transformer-based models, specifically focusing on source code syntax tokens and AST statements during both accurate and erroneous predictions. The empirical findings underscored an inclination of self-attention heads to allocate increased attention weights to select syntax tokens and AST statements during accurate predictions. Leveraging this insight, we introduced SyntaGuid, a mechanism designed to optimize the performance of fine-tuned PLMs in downstream tasks by directing self-attention heads toward specific source code elements. Empirical validations highlighted the efficacy of SyntaGuid, evidencing a marked enhancement in the performance of fine-tuned PLMs across diverse software engineering tasks, with an overall performance boost of up to 3.25\% and rectification of up to 28.30\% of previously erroneous predictions. We investigated on three software engineering tasks, a future direction could be evaluating the results on other software engineering tasks.

\section{Data Availability}
As part of our commitment to open science policy, all data collected for this study are made available as supplemental material. We provide our replication package in~\cite{AttentionBias}.